\documentclass[aps,twocolumn,showpacs,superscriptaddress,pra,10pt]{revtex4-1}
\usepackage{graphicx}
\usepackage{dcolumn}
\usepackage{bm}
\usepackage{color}
\usepackage{amsmath}
\usepackage{amsfonts}
\usepackage{subfigure}
\usepackage[makeroom]{cancel}
\usepackage{float}
\usepackage{xcolor}
\usepackage{ulem}

\newcommand{\COMMENTED}[1]{}

\begin{document}

\author{Annette Lopez}
\affiliation{Department of Physics, California State University Fresno, Fresno, California 93740}
\affiliation{Department of Physics, Brown University, Rhode Island 02912}
\author{Patrick Kelly}
\affiliation{Department of Physics, California State University Fresno, Fresno, California 93740}
\author{Kaelyn Dauer}
\affiliation{Department of Physics, California State University Fresno, Fresno, California 93740}
\author{Ettore Vitali}
\affiliation{Department of Physics, California State University Fresno, Fresno, California 93740}
\affiliation{Department of Physics, The College of William and Mary, Williamsburg, Virginia 23187}

\title{Fermionic Superfluidity: From Cold Atoms to Neutron Stars}
\begin{abstract}
From flow without dissipation of energy to the formation of vortices when placed within a rotating container, the superfluid state of matter has proven to be a very interesting physical phenomenon. Here we present the key mechanisms behind superfluidity in fermionic systems and apply our understanding to an exotic system found deep within the universe---the superfluid found deep within a neutron star.  A defining trait of a superfluid is the pairing gap, which the cooling curves of neutron stars depend on. The extreme conditions surrounding a neutron star prevent us from directly probing the superfluid's properties, however, we can experimentally realize conditions resembling the interior through the use of cold atoms prepared in a laboratory and simulated on a computer. 
Experimentalists are becoming increasingly adept at realizing cold atomic systems in the lab that mimic the behavior of  neutron stars and superconductors. In their turn, computational physicists are leveraging the power of supercomputers to simulate interacting atomic systems with unprecedented accuracy. This paper is intended to provide a pedagogical introduction to the underlying concepts and the possibility of using cold atoms as a tool that can help us make significant strides towards understanding exotic physical systems. 
\end{abstract}
\maketitle

\section{Introduction}
Moving from the ``comfort zone'' of classical mechanics to the more mysterious realm of quantum physics opens the possibility to explore the behavior of nature at a deeper and very fascinating level. In general, the transition from classical to quantum mechanics corresponds to the investigation of microscopic systems: we know that the motion of matter at the length scale of nuclei and electrons is governed by the celebrated Schr\"{o}dinger equation. Nevertheless, in some important cases, quantum effects do not remain hidden in the realm of microscopic systems but manifest in macroscopic systems, giving rise to plenty of exciting and puzzling physical behaviors that challenge our intuition and understanding. In fact, the subtle interplay between quantum mechanics, quantum statistics, and interatomic forces give rise to a huge number of fascinating phenomena. 

A crucial example of such a phenomenon is certainly superfluidity, together with the closely related superconductivity. Superfluidity is a unique state of matter characterized by the ability of a system to flow without friction. This results in some spectacular behaviors strongly challenging our common sense, like the observation of thin films of liquid helium climbing the walls of a container seemingly defying gravity. Moreover, the superfluid state of matter displays other unique properties, for example, the formation of quantum vortices when the system is enclosed in a rotating container. Although historically superfluidity was first observed in bosonic systems, and namely in $^4$He samples at temperatures below the celebrated ``lambda-temperature'', $T = 2.2$ K, we now know that the phenomenon can also be observed in fermionic systems. Even more interestingly, it appears to play a crucial role in some of the most fascinating and elusive systems in the universe, like unconventional superconductors and neutron stars. To our knowledge, superfluidity, and in particular fermionic superfluidity, is described only in very advanced books and papers, which makes it very hard for a student to have access to this very beautiful chapter in physics. We argue on the other hand that although the phenomenon appears, or is expected to appear, in some of the most complicated systems in nature, like quantum liquids, the interior of neutron stars and superconductors, the underlying physics can be captured by some simple and fundamental ``ingredients''. 

The purpose of this work is to present the key physical mechanisms underlying superfluidity in a simple but precise way, so that students or teachers will be able to comfortably fill the gap between basic quantum mechanical problems, like free particles, harmonic oscillators and simple atoms,  and more mysterious and fascinating systems, like superfluid neutrons deep inside a neutron star. We will also shed light into the connection between two apparently disconnected fields of physics, and namely atomic physics and nuclear astrophysics. The term ``universality'' frequently emerges in physics, and in this context it may be the key to ``reproduce'' in a laboratory on earth, at temperatures of the order of $10^{-9}$ K, the conditions that exist deep within a neutron star, where the temperature is of the order of hundreds of millions of Kelvins.
In a few words, if we put together a collection of fermions, for example $^6$Li atoms in a laboratory or neutrons in a star, and we have an attractive force acting among them, then we are in the conditions to observe a Fermi superfluid and the behavior of this superfluid will be universal, largely independent from the properties of the microscopic constituents. We will discuss the foundations and we will discuss the evidence that strongly suggests that such a system indeed exists inside a neutron star and it plays a major role in the behavior of the star itself.


This paper is organized as follows: we start with an introduction to Fermi gases, beginning with a description of non-interacting systems and moving to the ``universality" of an ultracold interacting Fermi gas, then apply our discussion to the proposed superfluid deep inside a neutron star. We aim to elaborate on the phenomena of glitches in a neutron star as well as to introduce the superfluid pairing gap, which the cooling curves of neutron stars are dependent upon.

\section{Fermi gases}
Three ``ingredients'' play a crucial role in fermionic superfluidity: quantum mechanics, Fermi statistics and an attractive force among the fermions. We will now examine these in some detail and we will discuss the emergence of superfluidity from the interplay among the three. 

From fundamental quantum mechanics, we know how to begin building the mathematical description of the motion of $N$ spin-$1/2$ fermions (for example electrons or neutrons) of mass $m$. We ultimately want to find the wave function of the system, $\Psi({\bf{r}}_1\sigma_1, \dots, {\bf{r}}_N\sigma_N, \, t)$, where ${\bf{r}}_i \in \mathbb{R}^3$ is the position variable, $\sigma_i = \uparrow, \downarrow$ is the spin variable and $t$ is time. The wave function satisfies Schr\"{o}dinger's equation:
\begin{equation}
i\hbar\frac{\partial}{\partial t} \Psi = \hat{H} \, \Psi
\end{equation}
where $\hat{H}$ is the hamiltonian operator of the system:
\begin{equation}
\begin{split}
& \hat{H} = \hat{T} + \hat{V} \\
& \hat{T} = - \frac{\hbar^2}{2m} \sum_{i=1}^{N} \nabla_i^2 \\
& \hat{V} = \sum_{i<j=1}^{N} v({\bf{\hat{r}}}_i - {\bf{\hat{r}}}_j) 
\label{interpot}
\end{split}
\end{equation}
and where $v({\bf{\hat{r}}}_i - {\bf{\hat{r}}}_j)$ is the interaction potential energy of particles $i$ and $j$. It is possible to add an external field in which our fermions move, like the electric field created by the nuclei in a solid, but we will not need these extensions for the purpose of this paper. For a more detailed discussion of this topic, please see for example \cite{PATHRIA2011231}. Since we are dealing with fermions, the wave function $\Psi$ is antisymmetric with respect to an exchange of particle labels, that is:
\begin{equation} \label{antisymfunc}
\Psi(\dots {\bf{r}}_i\sigma_i \dots  {\bf{r}}_j\sigma_j, \dots \, t) = - \Psi(\dots {\bf{r}}_j\sigma_j \dots  {\bf{r}}_i\sigma_i, \dots \, t) 
\end{equation}
This is a fundamental postulate of quantum mechanics. In fact, from (\ref{antisymfunc}), it follows that we can never have two fermions with the same spin orientation occupying the same position. 

\begin{figure}
	\begin{center}
	\includegraphics[trim = 30 30 30 0,clip,width=9.0cm]{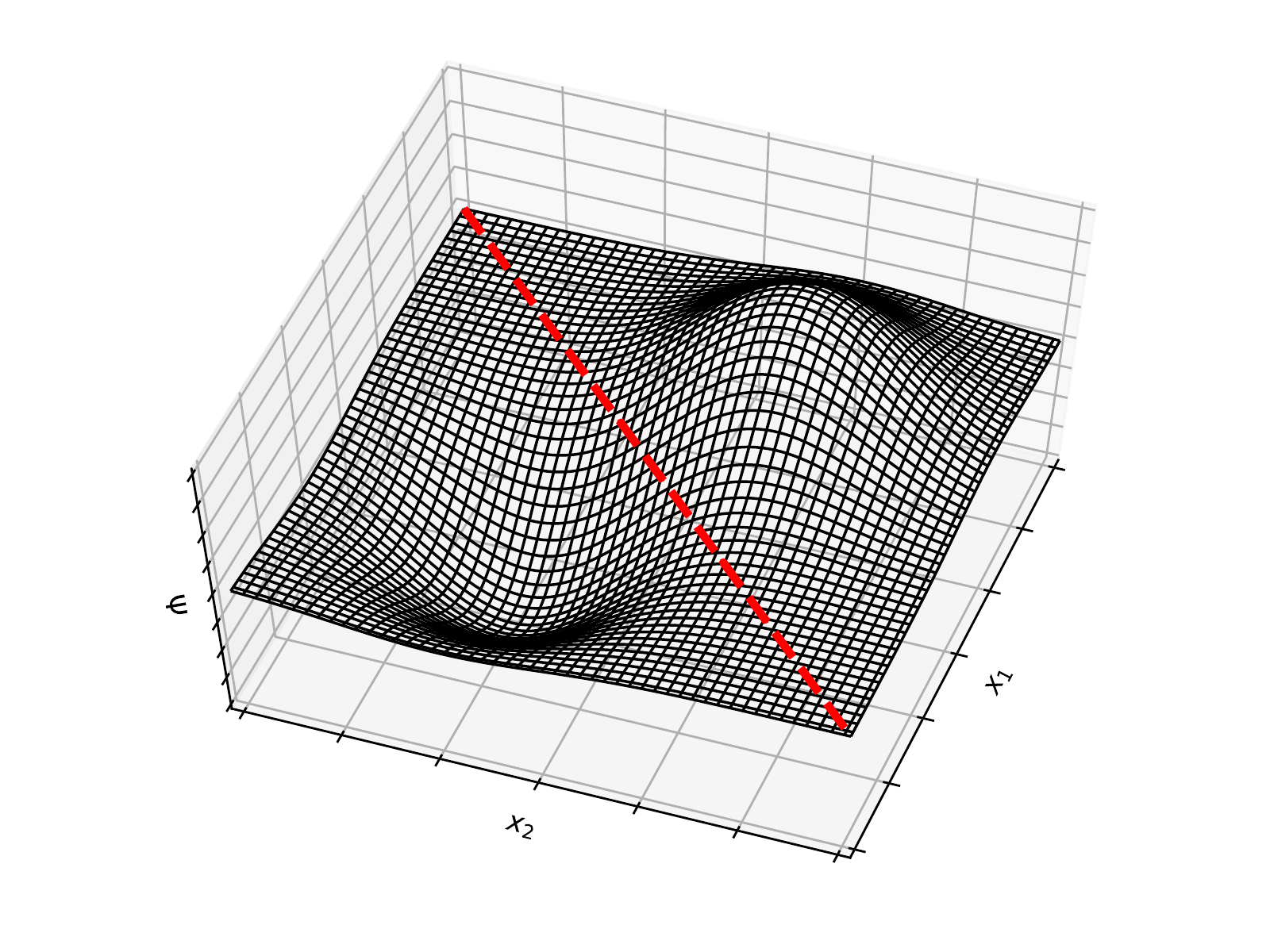}
	\caption{Illustration of fermion wave function antisymmetry.}
	\label{anti}
	\end{center}
\end{figure}

To understand this point better, imagine we have a source preparing two quantum particles, say two neutrons, with the same spin orientation $\sigma$, for example $\sigma=\uparrow$. For simplicity, we assume the particles move in one dimension.
The wave function of the system, $\Psi(x_1, x_2)$, where we omit the spin variables to keep the notation simple, yields the probability density for the positions of the two particles:
\begin{equation}
\left| \Psi(x_1, x_2) \right|^2 dx_1dx_2
\end{equation}
In other words, it is the probability to observe one neutron between $x_1$ and $x_1+dx_1$
and one neutron between $x_2$ and $x_2+dx_2$.
A fundamental principle in quantum statistics is that identical particles are indistinguishable: it is impossible to design an experiment which allows us to label the particles. This implies that the probability density must remain invariant if we swap the variables:
\begin{equation}
\left| \Psi(x_1, x_2) \right|^2 = \left| \Psi(x_2, x_1) \right|^2
\end{equation}
When the particle are fermions, in particular, the above condition is realized by antisymmetric wave functions:
\begin{equation}
 \Psi(x_1, x_2) = - \Psi(x_2, x_1)
\end{equation}
An example of an antisymmetric wave function is illustrated in figure \ref{anti}.
An immediate consequence is that the wave function identically vanishes whenever $x_1 = x_2$, the dotted line in the figure: two fermions with the same spin orientation cannot be in the same position. 
More generally, two fermions cannot be in the same quantum state: this is the mathematical expression of the celebrated Pauli principle.

\subsection{The Fermi sea}
Whenever there is no interaction and thus the hamiltonian has only the kinetic term $\hat{H}=\hat{T}$, it is relatively simple to solve Schr\"{o}dinger's equation. 
In particular, we can explicitly find the lowest energy stationary state of the system, called the ground state of the system. Mathematically, the ground state is the eigenvector of the hamiltonian corresponding to the minimum eigenvalue. Physically, the ground state wave function is the equilibrium state of the system when the temperature is $T=0$ K and it is frequently called the Fermi sea. In this section we will describe the Fermi sea, focusing on the physical interpretation.



As is traditionally done in condensed matter textbooks, we would like to take a moment to address the wave nature of particles. In the quantum regime, each fermion has a wave function to describe its motion. In the Fermi sea, the free fermions' wave functions take the form of plane waves. A plane wave has the mathematical expression $\varphi_{\bf{k}}({\bf{r}}) \propto \exp({i {\bf{k}} \cdot {\bf{r}}})$ and it describes a fermion with a well defined momentum $ {\bf{p}} = \hbar {\bf{k}}$ (${\bf{k}}$ is typically called wave vector)
and a well defined kinetic energy $\varepsilon({\bf{k}}) = \frac{\hbar^2 |{\bf{k}}|^2}{2m} = \frac{\bf{p}^2}{2m}$. Incidentally, we invite the reader to think of Heisenberg principle: as the momentum ${\bf{p}} $ can be measured with arbitrary precision when a fermion is in a plane wave state, the particle is completely delocalized: in fact, the probability density for the position of the fermion $|\varphi_{\bf{k}}({\bf{r}})|^2$ is independent from its position.
Due to Pauli Principle, each plane wave can accommodate up to two fermions, one with spin up ($\uparrow$) and one with spin down ($\downarrow$).  
We now set $\frac{\hbar^2}{2m} = 1$ for the rest of this discussion to focus on the concepts at hand. 

In simple words, each fermion has a well defined value of the momentum, or equivalently of the velocity\COMMENTED{: this resembles the well known classical Newton's law stating that the velocity of a particle will never change unless a net force acts on it}. In a classical context, at  $T=0$ K, all the particles would be ``frozen'', and all the velocities would be equal to zero. On the other hand, because of the Pauli principle, only two particles, one with spin up and one with spin down, are allowed to have zero velocity. Therefore, all of the other fermions must ``move''. \COMMENTED{More precisely, each plane wave can accommodate up to two fermions, one with spin $\uparrow$ and one with spin $\downarrow$.} 

From the mathematical point of view, it is convenient to enclose our fermions in a cubic box of volume $V$ and to fix the number of particles, say $N$. As it is traditionally done,
we choose simple periodic boundary conditions on the walls of the box: the idea is that the box is not a physical container, but it is a region inside a huge Fermi gas, where we are floating, and periodic boundary conditions allow us to mimic an infinite system made of an infinite number of replicas of our box. The volume $V$ induces a discretization in the momenta, as only plane waves whose wavelengths fit the box are allowed. This is analogous to a vibrating string with the two ends fixed: not all vibration modes are allowed.

\begin{figure}
	\includegraphics[trim = 50 50 50 50,clip,width=7.62cm]{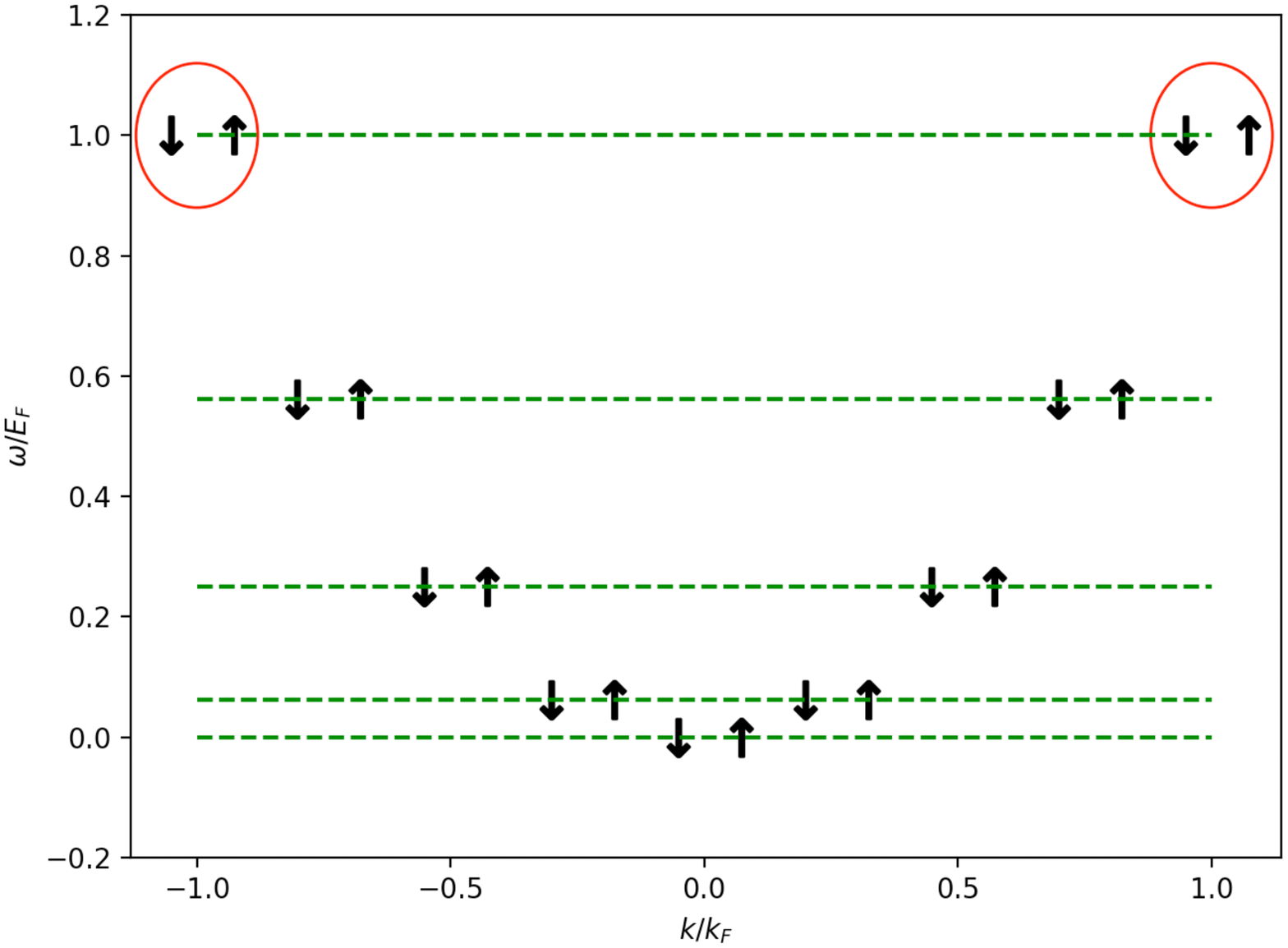}
	\caption{Illustration of a one-dimensional dispersion relation, showing also the fermions ``sitting'' on the available momenta.The green dashed lines are occupied energy levels and the red circles highlight the Fermi surface (in this case, two momenta).}
	\label{1D}
	\includegraphics[trim = 50 40 50 50,clip,width=7.62cm]{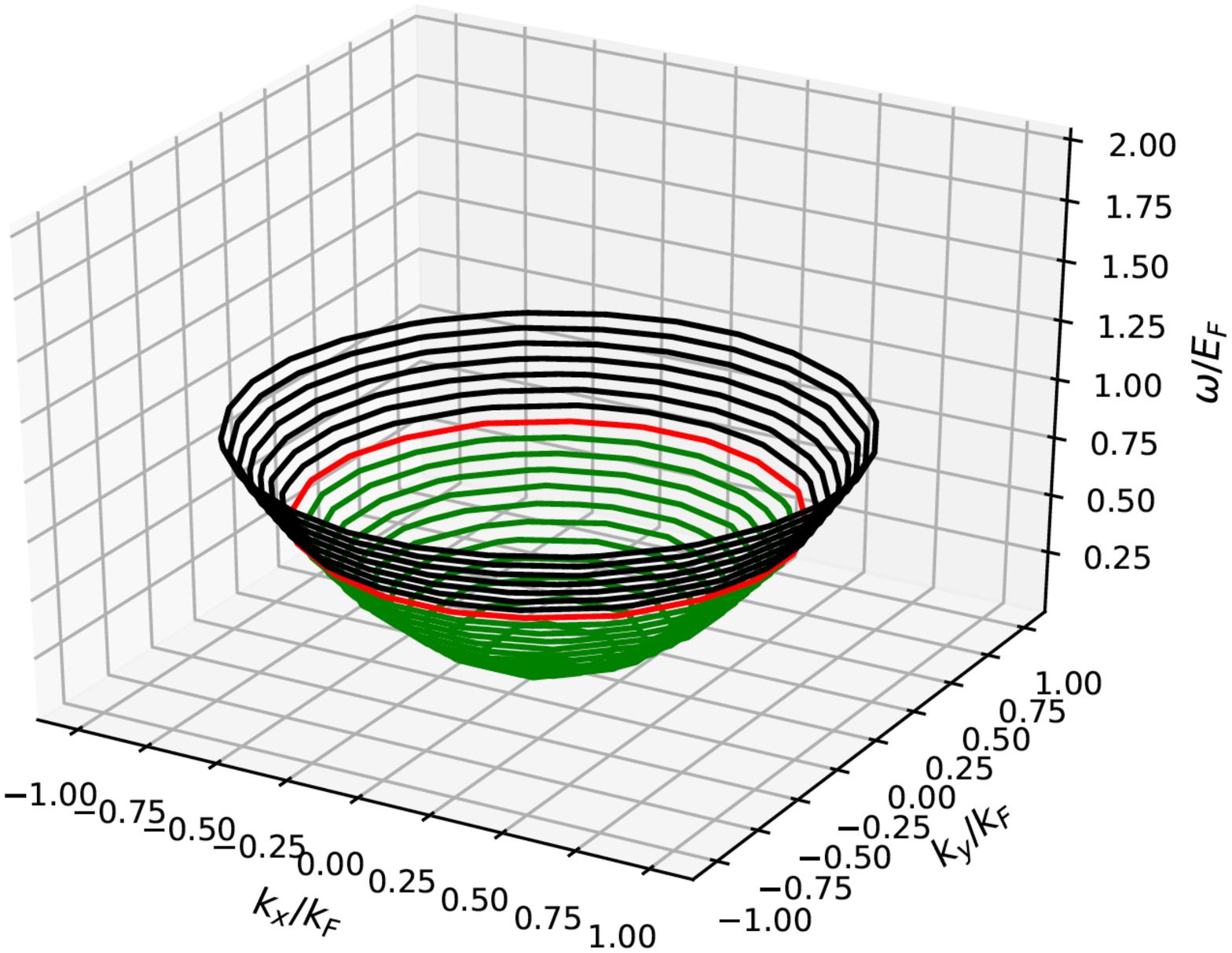}
	\caption{In two dimensions, the dispersion relation takes the form of a paraboloid. The red circle represents the Fermi surface, the green represents occupied energy levels, and the black represents unoccupied energy levels.}
	\label{2D}
\end{figure}

\begin{figure}
	\includegraphics[trim = 70 40 70 40,clip,width=7.62cm]{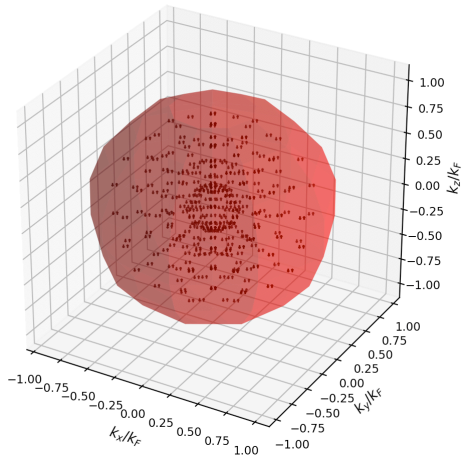}
	\caption{Illustration of a three-dimensional Fermi phere. The Fermi surface is the surface of the sphere. 
	}
	\label{3D}
\end{figure}

In the minimum energy state, the plane wave states will be filled starting from the lowest kinetic energy at ${\bf{k}}=0$ and 
continuing until all  $N$ fermions have been placed. All the plane waves corresponding to higher kinetic energies remain unoccupied. The Fermi sea in one and two dimensions can be illustrated as in figures \ref{1D} and \ref{2D}, where we show the fermions ``sitting'' on the available momenta. For each momentum, we will have one fermion with spin up ($\uparrow$) and one with spin down ($\downarrow$). It is useful to show such available momenta together with the related kinetic energies, often denoted as $\omega$: these plots, shown in figures \ref{1D} and \ref{2D} for one and two dimensional Fermi gases respectively, are known as dispersion curves.
Geometrically, the set of occupied momenta is a sphere in three dimensions 
, visualized in figure \ref{3D}: this sphere is called
the Fermi sphere and it fully characterizes the state of the fermions in equilibrium at T $= 0$ K. 
The radius of the Fermi sphere \COMMENTED{, if  we assume that the fermions move inside a cubic box of volume $V$,}  is given by:
\begin{equation}
\label{kfermi}
k_F = \Big(\frac{3\pi^2N}{V}\Big)^\frac{1}{3}
\end{equation}
and is called the Fermi momentum of the system. The surface of this sphere, which separates the occupied momenta from the non-occupied momenta, is called the Fermi surface. Associated with the Fermi momentum we have the Fermi energy, which is the maximum kinetic energy occupied by a fermion at T $= 0$ K, defined as
\begin{equation}
E_F = k_F^2
\end{equation}
Thus, fermions are considered to be within the Fermi sphere if 
\begin{equation}
{\bf{k}}^2= k_x^2+k_y^2+k_z^2< E_F
\end{equation}
is satisfied.
Intuitively, when we observe the system in momentum space, the distribution of fermions in the Fermi sea resembles a parking lot, which is completely  filled up to the Fermi surface, and empty outside the Fermi sphere. In this language, Pauli principle simply states that two cars cannot share the same parking spot. One spot corresponds to a value of ${\bf{k}}$ and one orientation of the spin. 

A consequence of this organization is the concept of degenerate Fermi pressure. The amount of ``space'' is limited in the parking lot, and once full, we can no longer accommodate new arriving cars. The system thus exerts a "pressure" on any newcomers trying to squeeze themselves in, barring any entrance into the Fermi sphere. Similarly, the system will resist any external forces trying to compact the Fermi surface: the parking lot size is fixed and unchanging. This structure of Fermi gases can be found in many different types of systems. The most well known Fermi gas that exists in nature is the fluid of electrons inside materials, the main character of solid state physics classes, which we will not discuss here. Instead, we aim to focus on the gas of neutrons within the core of a neutron star.

\subsection{How cold is a Fermi sea?}
It is very common in physics to transform energies into temperatures. In our context, the Fermi temperature of a system is defined as $T_F = \frac{E_F}{k_B}$, where $k_B$ is the usual Boltzmann's constant. We described above the minimum energy state of the system, that is, the state that the system chooses when thermodynamic equilibrium is established at $T = 0$ K. However, it can be seen that as long as the physical temperature remains much lower than the Fermi temperature of the system, $T << T_F$,  the temperature dependence of the
 macroscopic properties of the system, for example pressure and energy, is negligible and the system can be described as if it were at $T=0$ K. So, while classical systems immediately start to move as we heat them up starting from $T=0$ K, fermionic systems will just ``ignore'' the change in temperature until the temperature approaches the order of the Fermi temperature. Applying this concept can be very counter-intuitive when looking at ``natural" Fermi gases, such as neutron stars, since we typically consider them to be ``very hot". When describing the interior temperatures of neutron stars \cite{neutronstars}, which are on the order of $10^8$ K, we observe them to be well below their Fermi temperatures, $T_F \simeq 10^{12}$ K. This huge value is due to the density of the star: in fact, from \eqref{kfermi}, we see that the Fermi energy, and thus the Fermi temperature, depends on the density of the particles. Therefore, we can consider the interiors of these stars as ``very cold" even if their actual temperature is hundreds of millions of Kelvins! 

In the lab, we can artificially generate ultracold Fermi gases \cite{RevModPhys.80.1215} (also referred to as cold atoms) which are way less dense than a neutron star and have a Fermi temperature around $T_F \simeq 10^{-8}$ K. It is now possible to prepare these gases at temperatures of the order of $T \simeq 10^{-9}$ K, so that even these systems can be considered (somehow more naturally) ``very cold''.

In the next section, we will discover that the two systems, neutrons in a star and cold atoms, share even more properties, other than being both ``very cold'': they are both superfluids and their behavior may be surprisingly similar.

\COMMENTED{
\textcolor{red}{
We also observe that the Fermi energy depends on the density $N/V$ of the system: dense systems have high Fermi energies while dilute systems have small Fermi energies. Neutron stars can be thought of essentially as giant nuclei: the opposition to gravitational collapse is solely provided by degenerate Fermi pressure, a consequence of the neutrons Fermi sea organization (the parking lot is full, in our previous analogy). Neutron stars are the smallest, densest stars known with a typical mass of 1.5 solar masses and radii of about 10 km (the size of a city!). So why can we compare a system that can be made in a laboratory setting with giant stars in our universe? We must first introduce forces acting amongst the fermions, presented in the next section.}
}


 \subsection{Interactions and the origins of superfluidity: ``universality'' of Fermi gases} \label{dilute}
 We stress that the Fermi sea as we describe it above cannot be a superfluid: there is no way our collection of fermions, ``piled-up'' like cars in a full parking lot, can flow without friction, which is the defining property of a superfluid. In order to have a flow without friction, we need to have a mechanism that prevents the energy exchange between the system and the walls of the container in which the system is flowing. In the case of the Fermi sea, energy exchange is readily available: injecting energy into the Fermi sea simply promotes one or more fermions to some unoccupied momenta above the Fermi surface. So, there is no way a simple Fermi sea may be a superfluid: for this, we need to dig deeper by introducing interaction. 
 
The model of a non-interacting Fermi gas is extremely useful, and it helps us shed light into the microscopic and macroscopic behavior of plenty of physical systems.  But the picture becomes even more intriguing if we allow the fermions to interact at low temperatures, in particular when the interaction is attractive. 
The physical origin of the attraction depends on the details of the physical systems. For example, we know that the gas of electrons in metals experiences an attraction as a consequence of their interaction with the vibrating lattice of the positive charges (the nuclei of the metal atoms). While this attraction is normally negligible with respect to the Coulomb repulsion among the electrons, at very low temperature it becomes important, and it results in the phenomenon of superconductivity: electric current is able to flow without resistance; superconductivity is the direct analogue to superfluidity, they are two sides of the same coin. In cold atoms, the attraction is due to the effective interaction among the lowest energy states of the atoms which can be manipulated using external magnetic fields. The strong interactions among the neutrons, like those in the interior of a neutron star, is attractive in a given range of inter-particle distances. 

The detailed form of the interatomic potential from (\ref{interpot}), $v({\bf{r}}_i - {\bf{r}}_j)$, will naturally be specific to the physical system that we are investigating. In general, though, it will always have a range $r_0$ such that $v({\bf{r}}_i - {\bf{r}}_j) \simeq 0$ if $|{\bf{r}}_i - {\bf{r}}_j| > r_0$: two particles need to be close enough to feel the attractive force. For example, nuclear forces are by far the strongest forces that we are aware of, but (fortunately!) we never experience them in our everyday life since their range is as tiny as the size of an atomic nucleus.
 In most interesting situations involving Fermi gases, it happens that the average distance between two fermions, which can be defined as $d = \left(\frac{V}{N}\right)^{1/3}$, is much larger than the range $r_0$. This implies that the ``fine'' details of the interatomic force can be safely neglected. We say that the systems are ``dilute''. The possibility to ignore most of the details of the interatomic forces allows us to conclude that dilute gases are ``universal''. \COMMENTED{The specific details do not play a significant roles: they are made of fermions which attract each other, and this is all we need!} This allows us to look for similarities between cold atoms and the fluid of neutrons within a neutron star. In fact, both systems can be considered to be cold, as discussed above, and also to be dilute. In fact, it is known that in the inner crust of a neutron star the average distance among neutrons \cite{neutronstars} is much bigger than the range of nuclear forces, which is comparable to the size of nuclei, of the order of $10^{-15}$ m.
This means that the very complicated, and partly unknown, details of the force between two neutrons can be neglected, and the effect of the interaction can be taken into account by focusing on the relative motion of two neutrons when they are very far from each other. Scattering theory (please see e.g. \cite{griffiths_schroeter_2018}) dramatically simplifies the description of this ``dilute" regime: at low energy, only a few physical quantities are enough to capture the effect of the interatomic forces. The properties of dilute Fermi gases, from cold atoms to neutron stars, are ``universal'' in the sense that they depend only on the product $k_F a$ \cite{PhysRevC.77.032801}. Here, the Fermi momentum $k_F$ tells us how dense the system is while $a$ is the scattering length \cite{griffiths_schroeter_2018,RevModPhys.80.1215,PhysRevC.77.032801} which is a property of the relative motion of two particles and tells us how strong the interatomic forces are.\COMMENTED{, where $k_F$ is the Fermi momentum, which tells us how dense is the system, while $a$ is the scattering length which tells us how strong are the interatomic forces.}

The attractive forces may induce the fermions to organize in pairs: in the simplest cases, each spin-up fermion will ``find'' one spin-down partner, and they will bind to form a pair. The crucial point is that these pairs, composite states of two spin $1/2$ fermions, act as bosons, and the spectacular phenomenology related to Bose statistics enters the game. 
In particular, Bose-Einstein Condensation (BEC) may take place, leading a macroscopic number of pairs to populate a single quantum state \cite{PATHRIA2011179}. In a fermionic system each quantum state can accommodate at most one fermion. However, in a bosonic system, a macroscopic number of bosons will condense on the same quantum state when the temperature is very low. An instability takes place here; as fermions start to pair, the collection of bosonic molecules may condense, and a macroscopic number of them will be described by a single-particle (here more properly single-pair) quantum wave function. 

The system behaves in a {\it{coherent}} way \cite{PATHRIA2011179}: in very simple words, a huge number of pairs are doing exactly the same thing! In more precise terms, we have a macroscopic occupation of a quantum state, which thus manifests at the macroscopic level. Particles lose their individual identities as their wave functions overlap and begin to act in unison. This paves the way for superfluid behavior. In fact, in order for an energy exchange to happen between the system and its environment, a pair needs to be broken first. The energy needed to break a pair of fermions is called the {\it{superfluid gap}} of the system, often denoted as $\Delta$, and it is a crucial property of superfluids. 
The gap limits the amount of energy exchange the system has with its environment, therefore, this new minimum energy state is now ``protected'' and can flow without dissipation.

We stress that, when the superfluid is on earth,  $\Delta$ can be measured experimentally with spectroscopy experiments (please see for example \cite{schwabl}) and, from the theoretical point of view, its calculation is a big challenge. Recently, exact results have started to appear in computational studies of cold atomic Fermi systems \cite{PhysRevA.96.061601,PhysRevA.102.053324}. This energy gap has deep implications for the properties of the system. One physical quantity that is strongly affected is the specific heat, which can be shown to contain a term of the form $\exp(-\Delta/k_BT)$ \cite{pulsars}, and has important consequences for the thermodynamic properties of the system. In the realm of neutron stars, for example, the superfluid gap is expected to affect the cooling curves of the stars \cite{pulsars}, which describe how the temperature of the star evolves as a function of time after the star is born. Very interestingly, if we are able to find a good estimation of the pairing gap through the analysis of the cooling curves, then we may learn something about the superfluid that exists deep in the star, and thus having access to properties of matter in a very exotic environment. The ``universality'' of Fermi superfluids that we discussed earlier would then allow us to mimic these conditions in a cold gas, mimicking a system that exists in the deep universe.

 \subsection{The wavefunction of the coherent state}
In the superfluid state, since a very large number of the particles have the same behavior we do not need to solve for the wave function of a many particle system. We have a condensate and thus it is enough to solve for the wave function of a single boson. A percentage of the particles will be in a higher energy state when the system has a finite temperature, but for the sake of simplicity we ignore these particles if the temperature is sufficiently low. In order to keep the math as simple as possible, we will disregard the internal structure of a pair of fermions and will treat the pair just like a structureless boson. Therefore, the wave function governing the macroscopic motion of the superfluid will have the general form:
\begin{equation}
\label{polar}
\Psi({\bf{r}}) = R({\bf{r}}) e^{i \Theta({\bf{r}})} 
\end{equation}
where $\rho({\bf{r}})=R({\bf{r}})^2$ is the probability density for the position of the bosons, while the phase $\Theta({\bf{r}})$ controls the current density, that is, the local velocity of the particles. Note that (\ref{polar}) has no dependence on time. In general, the wave function is a function of time, however, here we have assumed the system has already relaxed to equilibrium. This equilibrium function is also known as a {\it{steady state}}. Once the steady state is reached, the state will not change, unless the system is somehow perturbed and driven out of equilibrium.
The wave function \eqref{polar} entirely governs the motion of a superfluid and in particular it allows us to understand the formation of quantum vortices, which is one of the crucial features of superfluids and are the main reason why it is widely believed that neutrons deep in a neutron star are superfluid. Before discussing this important topic, we will spend some time ``diving'' into a neutron star, where one of the most fascinating superfluids exists.

\section{The Formation of a Neutron Star}
Now that we have presented the key ingredients that allow us to understand the physical mechanisms underlying the spectacular phenomenology of superfluids, we find it a beneficial example for deepening our understanding to discuss one of the most mysterious superfluids that exists in nature: the fluid of neutrons in a neutron star. We will certainly need some context, so we will briefly provide a general description of the stars, starting from their birth. Later, we will discuss how the superfluid influences, or is speculated to influence, the behavior of the star through a description of glitches and the cooling process.

Interstellar gas clouds are diffuse and primarily composed of hydrogen molecules and dust (particles made of carbon, silicon, and oxygen atoms). These gaseous conglomerates can span distances from less than a light-year to several hundred light-years. 
We see the temperature vary inside the interstellar clouds: dust absorbs visible and ultraviolet light, cooling the interior regions. In denser regions, a lack of thermal pressure creates an instability, and a
perturbation to the cloud, such as a shock wave from a supernova or nearby colliding galaxies, can induce collapse. Under the influence of gravity, gas will fall to the center of mass. Gravitational potential energy is converted into heat energy under this compression, creating a thermal pressure gradient. If the core temperature reaches $1.5\cdot10^{7}$K, nuclear fusion jump-starts and the cloud becomes a ``protostar'', beginning its journey on the main sequence of stars.
Thermonuclear fusion sustains the star as thermal and radiation pressure balance gravity for millions to billions of years. If the mass of the star is big enough, the core will burn through a chain of elements: hydrogen, helium, carbon, neon, oxygen, magnesium, silicon, and iron. As each element is exhausted, the core contracts until the ignition temperature for the next step of the elemental chain is reached. 
When iron, the most stable nucleus, is burnt, a critical (and quite dramatic) stage is reached: there will not be a new fusion reaction. The thermonuclear energy is not able to compete against gravitational collapse, and the star caves-in on itself. The situation inside the core during this critical stage is really unique: the energy is such that electrons will combine with the protons to form neutrons and neutrinos, in a process called inverse $\beta$-decay. The neutrons will organize in a Fermi sea, which, as we discussed earlier, resists the compression of the core, making it ``stiff.". Infalling material will thus rebound, sending out a shock wave. This wave will stall a few hundred kilometers from the center as its energy is dissipated via neutrino losses and photodisintegration of nuclei. Free-falling material once supported by the core will collide with this stalled shock front to produce another shock wave that ejects all but the stiffened core in a supernovae explosion. As a residual of the explosion, a protoneutron star is born.
A protoneutron star has a temperature on the order of $10$ MeV, and escaping neutrinos will cool the star to an MeV or less during the first thousand years of life.

\section{The Structure of a Neutron Star and the neutron superfluid}
\begin{figure}
	\begin{center}
	\includegraphics[trim= 100 30 100 30 , clip, width=8.5cm]{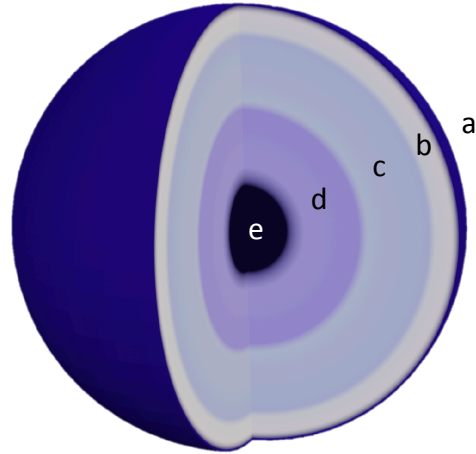}
	\caption{Structure of a neutron star: (a) the atmosphere and envelope are the outermost layer, followed by (b) the outer crust and (c) the inner crust, where the superfluid transition occurs, and (d) the outer core and (e) the inner core.}
	\label{star}
	\end{center}
\end{figure}

Glancing at a neutron star, one will immediately notice a hydrogen atmosphere. Directly beneath will lie a thermal insulator, known as the envelope, which acts as a barrier between the hot interior and the surface atmosphere. Diving deeper, we expect to see four internal regions of the neutron star that are distinguishable by nucleon densities: the outer crust, the inner crust, the outer core, and the inner core (depicted in figure \ref{star}). The outer crust is a few hundred meters below the envelope and is expected to be a nuclear lattice composed of heavy nuclei. The inner crust contains a lattice of neutron rich nuclei in equilibrium with a superfluid of neutrons which are paired in the $^1S_0$ quantum singlet state and a degenerate electron gas. Here the spectroscopic notation $^1S_0$ simply means that the neutrons in the inner crust form pairs made of one spin-up and one spin-down, like we have discussed in the previous sections.
 As we venture into the outer core, the system becomes more and more exotic: we do not have nuclei any more; instead, we have a fluid of free neutrons, protons, and electrons. It is assumed that both the neutrons and the protons are superfluids. The proton fluid is charged and would actually be considered a Type II superconductor. The neutrons would pair in a triplet state---two neutrons with the same spin orientation would pair---while the protons would form pairs in the more ``conventional'' $^1S_0$ singlet state.
A ``soup'' of two Fermi superfluids of protons and neutrons is thus expected to form the deepest region of the neutron stars: the interaction between the two components and the extreme density can make the physics very challenging. The inner core is mysterious because its composition and equation of state are unknown to the scientific community, however, there are several propositions for the composition of the core matter: nucleons; pion or kaon condensates; hyperons; and quarks. For the purpose of this paper, we focus on the superfluid of neutrons in the inner crust, which is ``dilute'' and universal, in the sense discussed in section \ref{dilute}; if we knew the interaction strength, or the superfluid gap, we would be able to ``observe'' it in a laboratory in the form of a cold atomic Fermi gas. Cold atoms would ``impersonate'' the neutrons and we would be able to tune the interaction strength to reproduce the physics inside the mysterious star. This may be achieved through observation of the cooling curves of the stars, which would allow us to infer the value of the interaction strength.
Before discussing this very important point, we would like to discuss the evidence for the existence of a neutron superfluid inside a neutron star: the phenomenon of glitches. \COMMENTED{ we have indirect evidence of the existence of vortices which, as we discussed earlier, are a crucial properties of a superfluid.}

\section{Vortices and Glitches}
\begin{figure}
	\begin{center}
	\includegraphics[trim = 80 20 70 20,clip,width=7.62cm]{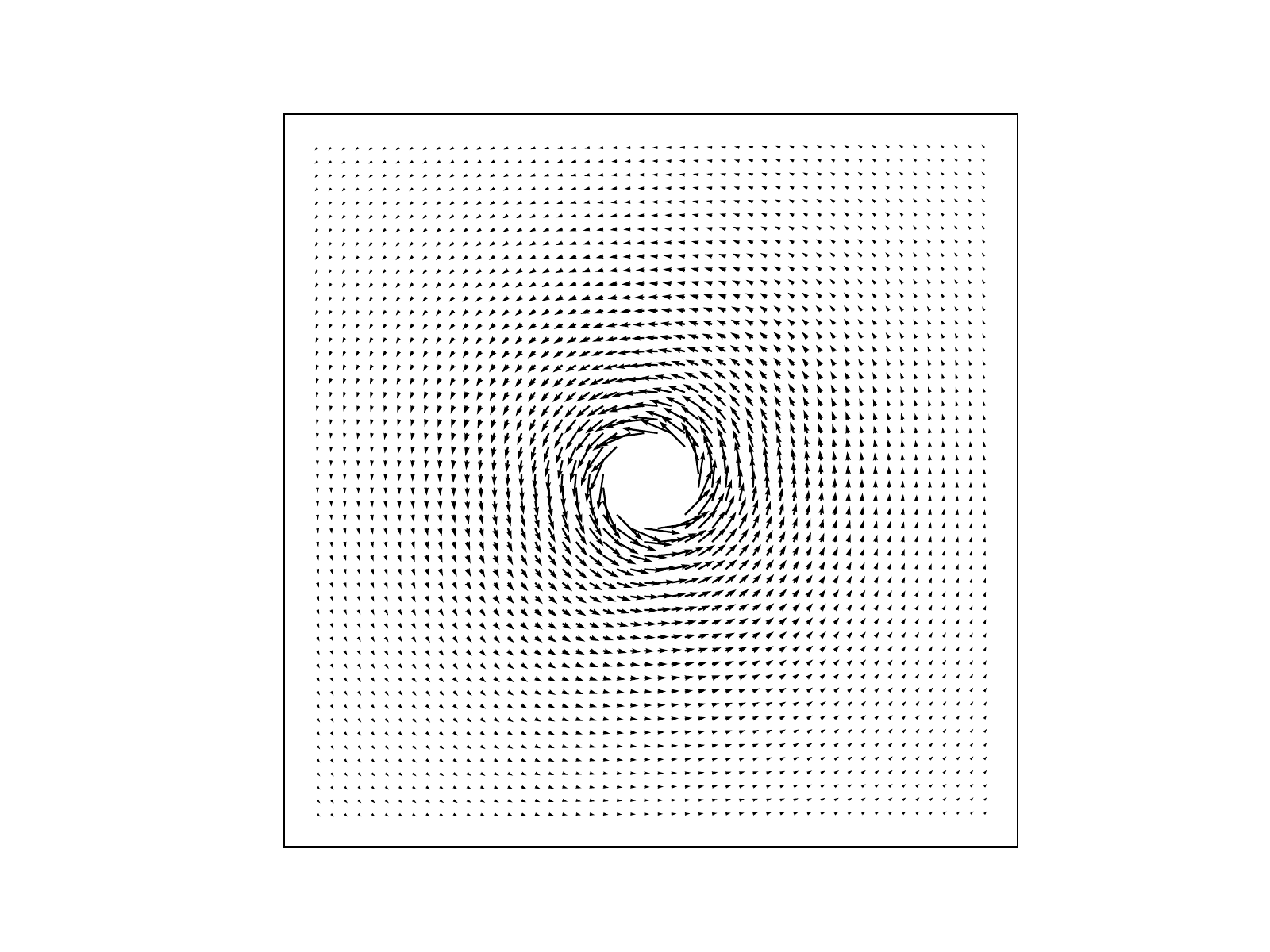}
	\caption{Top-view diagram of the vortex velocity vector flow direction field in \eqref{vortex}.}
	\label{vortexpic}
	\end{center}
\end{figure}

We now provide a short mathematical discussion of the superfluid vortices inside the rotating crust of a neutron star. As long as the rotation is slow, the superfluid will remain steady, since the container is not able to drag the nonviscous fluid. A general expression for the local velocity, given the wave function \eqref{polar}, is 
\begin{equation}
\label{velocity}
{\bf{v}}({\bf{r}})  =  \frac{\hbar}{m} {\nabla} \Theta({\bf{r}})
\end{equation}
which corresponds to the velocity field of the condensate.  We will derive this expression in the appendix.
For a steady or slowly rotating superfluid, the phase $\Theta({\bf{r}})$ does not depend on the position and thus we have no net velocity field.
The situation does change when the rotation becomes fast. The wave function that describes the motion of the system will correspond to the minimum energy when studied from a reference frame that is rotating with the container. Imagine, for simplicity, that the container is a cylinder that is rotating around its axis, which we chose to be the $z$ axis of our reference frame. It is natural, simply by transforming the energy into a rotating reference frame, to expect that the wave function $\psi({\bf{r}})$ will be an eigenvector of the component of the angular momentum along $z$, $\hat{L}_z$, implying that $\Theta({\bf{r}}) = m \varphi$, where $m$ is an integer number, while $\varphi$ is the cylindrical angular coordinate of the position ${\bf{r}}=(x,y,z)$: $\varphi = \arcsin(y/r)$, with $r=\sqrt{x^2 + y^2}$ being the distance from the axis. The relation \eqref{velocity} leads us to the interesting result:
\begin{equation}
\label{vortex}
{\bf{v}}({\bf{r}})  \propto  \frac{1}{r} {\bf{e}_{\varphi}} 
\end{equation}
where $ {\bf{e}_{\varphi}}$ is the unit vector along $\varphi$. This velocity field describes a vortex, as seen in figure \ref{vortexpic}. So, the fact that a superfluid is a coherent state of a macroscopic number of pairs described by a single-pair quantum wave function naturally allows us to conclude that, within a rotating container (provided that the rotation is fast enough), the superfluid will form vortices. From thermodynamic arguments it may be seen that, the higher the speed or rotation, the greater is the number of vortices. 

A detailed description of the origin of neutron star glitches would certainly require a highly sophisticated presentation, but we can capture the essence of the phenomenon with a few simple ingredients: a fluid of fermions (the neutrons) attract each other, form pairs, and as the star is rotating, form vortices. If the star is rotating (rotating neutron stars are typically called pulsars), the neutron superfluid inside a neutron star is expected to host an array of vortices. The fact that the superfluid coexists with a lattice of nuclei makes it possible for the vortices to be pinned to the lattice sites, similar to vortices in rivers that are pinned to rocks on the riverbed. The rotation of the star gradually slows down as the star emits energy. As this happens, the vortices will feel a force, known as a Magnus force, that pulls them towards the surface of the star. This force can break the pinning, and the vortex may hit the surface. As this happens, a glitch takes place---the star spins faster for a short time---and observations of these glitches give indirect evidence of a superfluid core. It is quite amazing to see how the simple physical mechanisms discussed in this paper allow us to give a possible explanation to a phenomenon that happens in the deep universe. 
We comment that other models have been proposed to explain the glitches, for example star-quakes, but to our knowledge the role of the superfluid vortices and their unpinning is widely accepted in the astrophysics community as the leading mechanism.

\section{Cooling of Neutron Stars}
While the glitches are certainly the most spectacular manifestation of the superfluid existing deep inside a neutron stars, the superfluid neutrons are expected to affect also other properties of the star, and in particular the cooling process.
A detailed description can be found in \cite{Shapiro}, and we just give a brief and non-exhaustive description here, in order to allow the reader to understand the role played by the superfluid.

Neutron stars undergo three main cooling stages. In stage one, the crust is thermally decoupled from the interior such that the surface temperature of the star reflects the thermal state of the crust. In the early life of a neutron star, $10-100$ years, cooling primarily happens via neutrino emission. During stage two, the surface temperature of the crust will adjust to the internal temperature, such that the core and crust reach thermal equilibrium. Within the period of $10-100$ years $ <  t < 10^5-10^6$ years, neutrino emission continues to dominate. Stage three refers to a mature neutron star experiencing photon emission via the transport of heat from the core to its surface. From $10^5-10^6$ years and onward, the evolution of the core temperature is thus governed by this radiation. The surface temperature of a star is dependent on the age of the star, and this relationship is shown with cooling curves.  

Cooling curves depend strongly on the properties of matter within the core of the neutron star, and are heavily dependent on the local density in the interior of the star and, in particular, they are sensitive to the existence of a superfluid. For a non-superfluid core, the transition from stage one to stage two consists of a sharp drop off in temperature due to the direct URCA process. A direct URCA process consists of a pair of reactions:
\begin{equation}
\begin{split}
& n \rightarrow p + e^- + \overline{\nu_e} \\
& p +e^- \rightarrow n + \nu_e
\end{split}
\end{equation}
where $n$ is a neutron, $p$ is a proton, $e^-$ is an electron, and $\nu_e$ is an electron neutrino. The emitted neutrinos are responsible for the cooling of the star, as they literally transport energy away from the star.
These reactions are possible given the right condition of density, $\rho \geq 4.62 {\rho_{nuclear}}$,  where the nuclear density is $ {\rho_{nuclear}} = 2.3 \times 10^{17} kg/m^3$ which is possible for stars whose mass is $M_{NS} \geq 1.35M_{\odot}$, where $M_{\odot}$ is the mass of the sun. This is because the resource of proton concentration must be sufficient (a ratio of the number of protons to the number of nucleons needs to be approximately $0.1$). 

In contrast, superfluidity suppresses the neutrino luminosity, slowing down the cooling in stage two. This is due to the fact that nucleons must be excited above the pairing gap (discussed in section \ref{dilute}) inherent in a superfluid to participate in the direct Urca process. A superfluid core will cool the star in stage one and stage two such that the observable crust temperature is lower than the critical temperature necessary for superfluidity. Because of this, neutron stars are ideal for determining the critical temperatures of superfluid matter at supernuclear densities.
Interestingly, an accurate analysis of the cooling curve may yield an estimate of the superfluid gap $\Delta$, which, in turn, may give us information about the Fermi superfluid, and namely the interaction strength. This information, thanks to the universality of Fermi superfluids, could allow us to mimic the neutron superfluid using cold atoms, as we will discuss in the next section.

\section{Experimental Analysis of the Superfluid Pairing Gap of a Neutron Star}
In the last few decades, after the celebrated realization of Bose-Einstein-Condensation in $1995$, the possibility of cooling down collections of atoms to temperatures very close to the absolute zero
opened the unique possibility to produce Fermi (and Bose) gases in a laboratory.  In the study of fermions, typically Lithium or Potassium atoms are cooled down with advanced cooling and trapping techniques. 
The resulting collections of very cold particles are realizations of quantum gases which can be studied with unprecedented experimental control. For example, we can control the interatomic forces by simply tuning an external magnetic field using a phenomenon known as Feshbach resonance. We are literally able to engineer a superfluid with a given value of $k_Fa$, as discussed above.

\begin{figure}
	\begin{center}
	\includegraphics[trim = 30 50 30 50,clip,width=7.62cm]{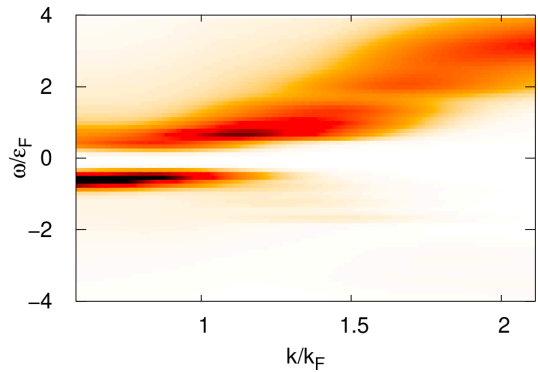}
	\caption{Color plot of the spectral function of a dilute attractive Fermi gas. Source: Annette Lopez and Ettore Vitali}
	\label{spec func}
	\end{center}
\end{figure}

For such ``artificial superfluids'', we can access the superfluid gap $\Delta$ and we can study its dependence on $k_Fa$, that is on the density and on the strength of the interaction. This can be done both experimentally and theoretically,  thanks to the unprecedented progress in computational methodologies that we witnessed in the last few decades.
From the experimental point of view, the study of the pairing gap relies on spectroscopy experiments, which measure the so called spectral function $A(\bf{k},\omega)$, yielding
a map of the quantum states that are available for the particles in the system.
The same function can be also estimated using theoretical and computational techniques,
and we show an example in figure \ref{spec func}. The zero of the energy is at the Fermi surface, or more precisely at the chemical potential corresponding to the given particle density. 
For $\omega < 0$, the spectral function tells us the probability that a particle with a certain momentum $\bf{k}$ has an energy $\omega$ (here, in units of the Fermi momentum and Fermi energy). The broadening is the corresponding uncertainty, arising from the fact that the momentum of one particle is not a good quantum number for an interacting system: a particle with a given momentum $\bf{k}$  can have an energy that is distributed around a mean. 
For $\omega > 0$, the spectral function shows us the available states for new particle that we may inject into the system in a spectroscopy experiment. From figure \ref{spec func} we clearly see a gap between the ``occupied'' states at $\omega<0$ and the ``available'' states at $\omega > 0$: this is exactly the superfluid gap.

\begin{figure}
	\includegraphics[trim = 30 -10 30 -20,clip,width=8.5cm]{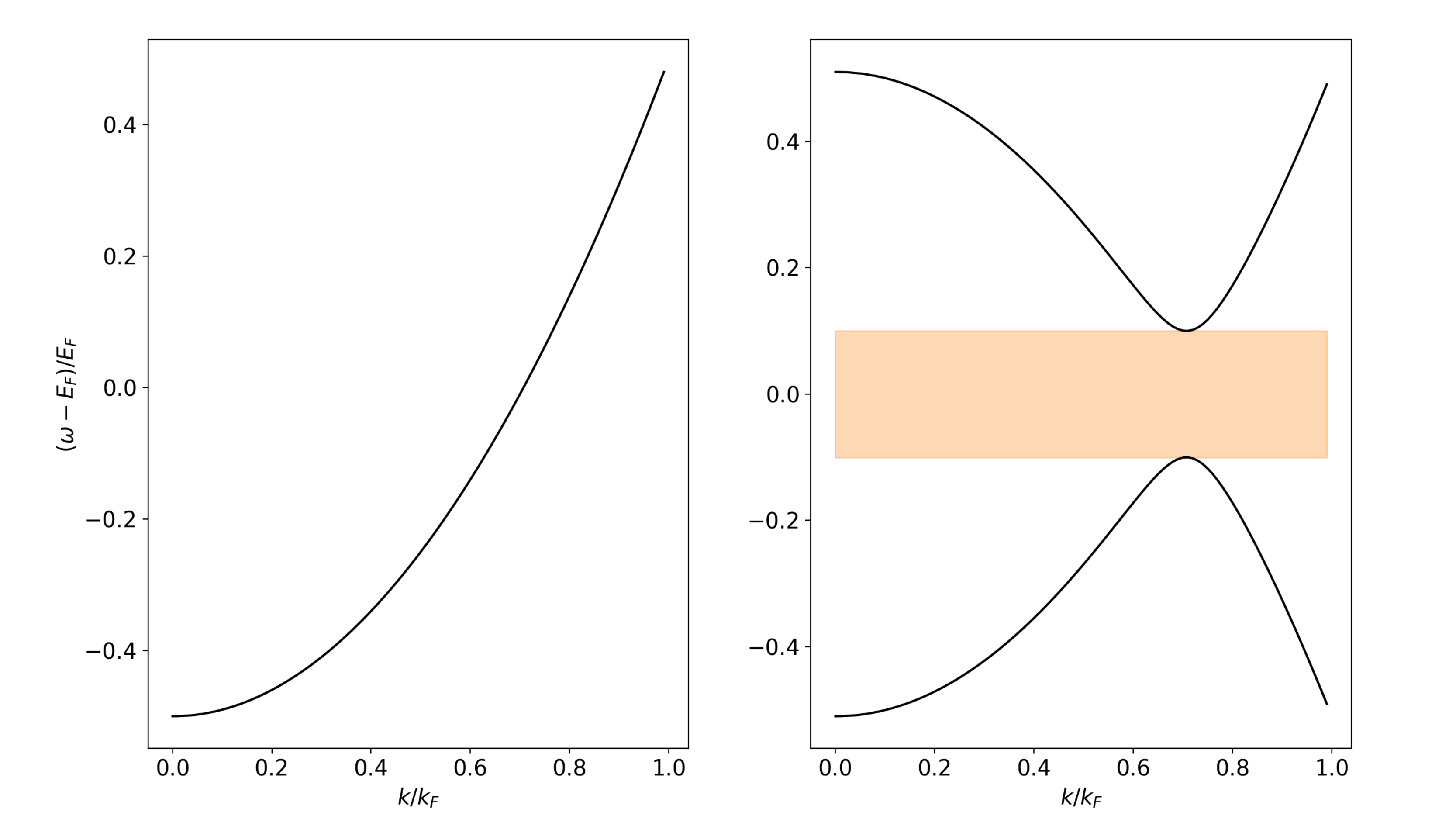}
	\caption{Left: dispersion curve for the non-interacting system. Right: dispersion curve of an interacting system. The interacting system has a prohibited region where no fermions may occupy energy states known as the superfluid pairing gap, $\Delta$ (highlighted in orange is $2\Delta$ symmetric about $0$ energy).}
	\label{example dispersions}
\end{figure}
The function $\omega(\bf{k})$, giving the mean energy as a function of the momentum and capturing the peaks in $A(\bf{k},\omega)$, is called the dispersion relation of the system.
In figure \ref{example dispersions}  we plot the dispersion relations for both a non-interacting Fermi gas, and for an attractive Fermi gas. In both cases the zero of the energy is set to the Fermi energy (or chemical potential for the interacting system), so that negative $\omega$ corresponds to occupied states and positive $\omega$ corresponds to unoccupied states.
In the non-interacting case, no energy is expended for a particle to move into a higher energy state. If we juxtapose the non-interacting dispersion curve with an interacting dispersion curve, shown in figure \ref{example dispersions}, we can clearly see the presence of a gap, visible around zero energy. In the interacting case, the dispersion curves follow the approximate relationship
\begin{equation}
\omega = \pm \sqrt{({\bf{k}}^2 - E_F)^2 + \Delta^2}
\end{equation}
where $\Delta$ is the gap. An interacting system has two branches, separated by the energy gap required to break apart a pair of fermions.  This gap is the superfluid gap of the system: the gas can flow without dissipation as long as the the container is unable to break the pairs. 
With cold atoms, like Lithium or Potassium, we can study these superfluids on earth and learn the dependence of the superfluid gas, and of other properties of the superfluid, like the vortices, on the interatomic forces, which can be controlled with high accuracy. Also, we can sharpen our theoretical and computational approaches to give more and more accurate predictions. Very interestingly, these systems are somehow universal, as discussed above, and so what we learn in a cold atomic system can tell us something about the superfluid neutrons deep inside a neutron star. A very exciting research direction is as follows: if we can infer the value of the superfluid gap from the cooling curves, as discussed above, can we learn more about the interaction strength and, maybe, ``reproduce'' the superfluid in a laboratory on earth?


\section{Summary and Conclusions}

Using quantum mechanics we can describe the nature of Fermi gases. Non-interacting fermions are fully governed by the Pauli exclusion principle which determines the organization of the Fermi sea: fermions will occupy all energy states up to the density dependent Fermi energy. In the quantum realm, each particle is in a plane wave, with a given momentum. Looking at a non-interacting Fermi gas in momentum space, we see that if any amount of energy is introduced to the system, fermions can easily use it to occupy states above the Fermi energy. 
Fermi gases have internal temperature scales dependent upon the Fermi temperature (directly related to the Fermi energy). So long as the Fermi gas's temperature is lower than the Fermi temperature, the system can be described as if held at absolute zero. As quantum effects become negligible when the temperature reaches the Fermi temperature, we can say that all quantum Fermi gases are ``cold''. 
If we allow the fermions to interact, if the average distance between the the fermions is much larger than the range of the force resulting from the interparticle potential, we can also declare a Fermi gas as being ``dilute''. 
Cold and dilute Fermi gases are ``universal''. This "universality" ultimately allows us to compare and connect two systems literally lightyears apart from each other: cold atoms made in a laboratory and the superfluid cores of neutron stars.

An attractive interacting potential amongst the fermions will result in pairs of fermions, which ultimately behave as bosons. These integer spin pairs no longer obey the Pauli exclusion principle and ultimately all collapse into the ground state, leading to a macroscopic manifestation of a quantum phenomenon---superfluidity---in which the pairs of fermions behave ``coherently''. 
In simple words, they all do the same thing! In more rigorous terms, a macroscopic number of fermions is described by the same single-pair ``collective'' wave function. A characteristic trait of a superfluid is the pairing gap: the energy required to break a pair of fermions. The energy states above the fermi energy become prohibited unless energy greater than or equal to twice the pairing gap is injected into the system. This is what allows a superfluid to flow without friction. Another characteristic trait is the formation of quantum vortices, when the system is in a rotating vessel, like the crust of a spinning neutron star.

Observations of glitches provides indirect evidence of the existence of a superfluid in the deep universe: a rapid acceleration in the rotation rate of a neutron star. These are predicted to be a result of the quantized vortices which can form within a rotating superfluid. Theorists have also concluded that the cooling of neutron stars, from birth to death, have a dependence on the pairing gap of the superfluid core. Ultracold atomic Fermi systems are a tool we can utilize to compute the pairing gap of a superfluid, which can be compared to models of cooling curves, shedding light on the interior of neutron stars.

Science is ultimately about observations and experiments on {\it{real}} physical systems. Experiments tend to be expensive and difficult, and thus the number of experiments and observations a researcher would like to do is much smaller than the number of experiments they actually can do. In addition, for practical reasons many systems in nature are impossible to study directly, such as most astronomical phenomena. Obviously the era of space exploration has allowed for us to have direct contact with the greater universe, but for the foreseeable future we are going to be confined within the solar system. The nearest neutron star is approximately 400 light years away, roughly 25 million times the distance between the Earth and the Sun. The technical challenge of recreating certain systems in the laboratory is also a major impediment to experimental science. Rapid progress is being made in realizing superfluid and supersolid phases in cold atomic gas systems. Because of the universality described above, these may serve as useful real models of the superfluid interiors of neutron stars.

We want to stress the importance of computational physics for investigating systems such as quantum fluids. Purely theoretical investigations are extremely difficult for complex systems. Computational physics can be seen as an implementation of theoretical physics at a much larger scale than can be done with pencil and paper work alone. Furthermore many of the methods implemented by computational physicists were invented before modern computers existed or were powerful enough to tackle physically meaningful systems. Computational physics will not be a replacement for experimental or theoretical work. However it combines the inexpensiveness of theory with the ability to account for the greater complexity of real physical systems. Computational results also serve as useful benchmarks that provide a roadmap for more ambitious experiments within the next decade.

\appendix

\section{Derivation and Motivation for the Probability Current}
\label{appA}

\indent Here we would like to provide a motivation for (\ref{velocity}). The explanation is straight forward but without it, this equation can seem to have appeared from nowhere. At first sight, the exclusive dependence on the phase can be quite perplexing. To understand this equation will start by deriving an expression for the probability current, and from there find (\ref{velocity}).
Suppose we have a region of space $\mathcal{V}$ and a particle with wave function $\Psi(\bold{r},t)$. We would like to know how the probability of finding the particle in $\mathcal{V}$ changes with time. We write the probability as
\begin{equation}
\label{totprob}
P(\mathcal{V})=\int_{\mathcal{V}} \rho \, d^3\bold{r}
\end{equation}
Where the probability density is $\rho = R^2 = |\Psi|^2 = \Psi^*\Psi$. Taking the time derivative of the density we have
\begin{equation}
\label{tprob}
\frac{\partial \rho}{\partial t}=\frac{ \partial \Psi^*}{\partial t}  \Psi + \Psi^* \frac{ \partial \Psi}{\partial t} 
\end{equation}
Substituting the Schr\"{o}dinger equation (where we assume we are working in units where $\hbar=1$)
\begin{equation} 
i \frac{\partial \Psi}{\partial t} = -\frac{1}{2m} \nabla^2  \Psi + V \Psi
\end{equation}
(and its complex conjugate) into (\ref{tprob}) we obtain
\begin{equation}
\label{divcur}
\frac{\partial \rho}{\partial t}=-\frac{i}{2m} (\Psi \nabla^2 \Psi^* - \Psi^* \nabla^2 \Psi) = -\nabla \cdot \bold{J} 
\end{equation}
Here we have defined a new quantity, the {\it{probability current}}
\begin{equation}
\label{current}
\bold{J} = \frac{i}{2m} (\Psi \nabla \Psi^* - \Psi^* \nabla \Psi)
\end{equation}
Using  (\ref{totprob}) and (\ref{divcur}) and applying the divergence theorem, we can write
\begin{equation}
\int_{\mathcal{V}} \nabla \cdot \bold{J} d^3\bold{r} = \int_{\partial{\mathcal{V}}} \bold{J} \cdot \bold{n} d^3\bold{r} 
\end{equation}
Where $\partial{\mathcal{V}}$ is the surface surrounding the region $\mathcal{V}$. 
\indent Intuitively, this means we can treat the probability density like an inhomogeneous fluid, that is, a fluid with a variable density, like a gas. As time passes in some regions of space probability can become more rarefied and in other regions probability can accumulate. The probability is a conserved quantity (the total amount is always equal to one) and obeys the continuity equation 
\begin{equation}
\frac{\partial \rho}{\partial t} = - \nabla \cdot \bold{J}
\end{equation}
which intuitively means that if a fluid moves from one region to another, it must move through all the regions in-between. This means any concentration of probability in one region will be at the expense of probability flowing out of adjacent regions.

We now want to obtain (\ref{velocity}). We start by writing the wave function in polar form using (\ref{polar}) and substituting this into the probability current given in (\ref{current}). This gives
\begin{equation}
\bold{J} = \frac{i}{2m} \big[R e^{i \Theta} \nabla R e^{-i \Theta} - R e^{-i \Theta} \nabla R e^{i \Theta}\big]
\end{equation}
Where we have suppressed the $\vec{q}$ and $t$ dependence.
Evaluating the gradients and simplifying we obtain
\begin{equation}
\label{finalcur}
\bold{J} = \frac{\rho}{m} \nabla \Theta(\bold{r},t)
\end{equation}
In classical hydrodynamics the current is given as $\bold{J}=\rho \bold{v}$. Here $\rho$ is the density of a fluid parcel moving with velocity $\bold{v}$. We can rewrite this expression in terms of momentum $\bold{p}$ as
\begin{equation}
\label{jandp}
\bold{J}= \frac{\rho}{m}\bold{p}
\end{equation}
Equating (\ref{finalcur}) and (\ref{jandp}) we have
\begin{equation}
\bold{v}(\bold{r},t)=\frac{1}{m}\nabla \Theta(\bold{r},t)
\end{equation}
Which gives us (\ref{velocity}): the velocity field of a particle in the superfluid.

%


\COMMENTED{
\newpage
\textbf{References}
\newline Beale, P. D. and Pathria, R. K. (2011). Statistical Mechanics.
\newline Gusakov, M.E. (2002). Neutrino Emission from Superfluid Neutron-Star Cores: Various Types of Neutron Pairing.
\newline Jahan-Miri, M. (2001). Glitches Induced by the Core Superfluid in a Neutron Star.
\newline Levenfish K. P. and Yakovlev, D. G. (1993). Suppression of Neutrino Energy Losses in Reactions of Direct Urca Processes by Superfluidity in Neutron Star Nuclei.
\newline Page, D. 2012. Pairing and the Cooling of Neutron Stars
\newline Page, D. and Applegate, J. (1992). Cooling of Neutron Stars by the Direct Urca Process.
\newline Page, D. and Reddy, S. (2006). Dense Matter in Compact Stars: Theoretical Developments and Observational Constraints.
\newline Schmitt, A. (2014). Introduction to Superfluidity.
\newline Seveso, S. (2014) Advances in Models of Pulsar Glitches.
\newline Thouless, D. J. et al. (n.d.) Quantized Vortices in Superfluids and Superconductors.
\newline Yakovlev, D.G. and Pethick, C.J. (2004). Neutron Star Cooling.
}

\end{document}